\documentclass[twocolumn,english,prl,showpacs]{revtex4}
\usepackage[T1]{fontenc}
\usepackage{textcomp}
\usepackage{amsmath}
\usepackage{amssymb}
\usepackage{graphicx}
\usepackage{setspace}
\usepackage{esint}
\usepackage{subfigure}

\makeatletter

\DeclareFontEncoding{LGR}{}{}
\DeclareTextSymbol{\~}{LGR}{126}

\@ifundefined{textcolor}{}
{
\definecolor{BLACK}{gray}{0}
 \definecolor{WHITE}{gray}{1}
 \definecolor{RED}{rgb}{1,0,0}
 \definecolor{GREEN}{rgb}{0,1,0}
 \definecolor{BLUE}{rgb}{0,0,1}
 \definecolor{CYAN}{cmyk}{1,0,0,0}
 \definecolor{MAGENTA}{cmyk}{0,1,0,0}
 \definecolor{YELLOW}{cmyk}{0,0,1,0}
 }

\makeatother

\usepackage{babel}
\begin{document}

\title{Ion acceleration in "dragging field" of a light-pressure-driven piston}

\author{Liangliang Ji\textsuperscript{1,2}, Alexander Pukhov\textsuperscript{1},
and Baifei Shen\textsuperscript{2}}

\address{\textit{\small \textsuperscript{1}Institut fur Theoretische Physik
I, Heinrich-Heine-Universitat Dusseldorf, 40225 Dusseldorf, Germany}}

\address{\textit{\small \textsuperscript{2}Shanghai Institute of Optics and
Fine Mechanics, Chinese Academy of Sciences, Shanghai 201800, China} }
\begin{abstract}
We propose a new acceleration scheme that combines shock wave acceleration (SWA) and light pressure acceleration
(LPA). When a thin foil driven by light pressure of an ultra-intense 
laser pulse propagates in underdense background plasma, it serves
as a shock-like piston, trapping and reflecting background protons
to ultra-high energies. Unlike in SWA, the piston velocity is not
limited by the Mach number and can be highly relativistic. Background protons can be trapped and reflected forward by the
enormous "dragging field" potential behind the piston which is
not employed in LPA. Our one- and two-dimensional particle-in-cell simulations and analytical model both show that
proton  energies of several tens to hundreds of GeV can
be obtained, while the achievable energy in simple LPA is below 10 GeV.
\end{abstract}


\maketitle
Laser-driven ion acceleration attracts more and more attention nowadays
because of its potential to realize table-top size ion accelerators,
which would greatly reduce the expense and occupying area compared
to conventional accelerators. In most proposals,
intense/ultra-intense laser pulses are employed to stimulate 
strong electrostatic fields with amplitude several magnitudes higher
than that of conventional methods. These fields move at high velocities thus ions can be continually accelerated to
large energy in a short distance. In Target Normal Sheath Acceleration
(TNSA) \cite{Roth2001-1,pukhov2001-1,Wilks2001,Snavely2000,Mora2003,Esirkepov2002,Schwoerer2006,Gaillard2011},
the sheath field, formed at the target back by laser heated super-thermal
electrons, moves at several to tens of the ion acoustic speed. Ions
co-move with the field and are accelerated to tens of MeV, while energy
above hundred MeV is quite a challenge because of the weak scaling
law versus laser intensity. In Light Pressure Acceleration (LPA) \cite{macchi2005,shen2001,esirkepov2004,Yan2008,Pegsorarol2007,Robinson2008,Henig2009,Chen2009},
electrons in the
thin plasma foil are pushed inward by the laser pressure and build
up an intense charge separation field, by which protons are accelerated.
As the foil is driven as a whole, protons move
along with the electrostatic field and are successively accelerated
to energies that theoretically may reach GeV. 
Nevertheless, even in simulations it is difficult to overcome the ten
GeV barrier for protons with LPA, since the accelerating gradient
drops quickly due to the relativistic Doppler effect. As soon as the
protons become relativistic, their energy grows very slowly with time ($\sim t^{1/3}$).

Recently, a breakthrough in collisionless shock wave acceleration
\cite{Denavit1992,Palmer2011,Habergerger2012,Silva2004,Forshund1970,Romagnanil2008}
has been reported \cite{Habergerger2012}. It suggests that a multi-pulsed
picosecond (ps) laser pulse can efficiently excite a collisionless
shock and maintain its velocity in a density-decaying plasma target.
The produced protons have much larger energy and lower energy spread
than predicted by the hole-boring shock \cite{Silva2004}. When the
shock propagates like a piston, some protons are trapped and
reflected to a speed determined by the energy density gradient. The exponentially decaying density profile of the
target is important for keeping the shock velocity high and constant.
However, plasma itself also expands at the ion acoustic speed simultaneously.
It seems inevitable that the pressure gradient will be further weakened
at a later stage. This would put a limit on the final energy gain.
From other side, the shock wave velocity is limited by the Mach number,
which cannot be very high \cite{Forshund1970}.

Generally, the shock front may be considered as a fast-moving
piston carrying a strong electrostatic field. Imagine, one could free
the limitation on its velocity, e.g., the shock becomes relativistic
\cite{relativistic shock}. Then, the trapped and reflected protons
would reach a relativistic $\gamma-$factor of about $\gamma=2\gamma{}_{s}^{2}$,
where $\gamma{}_{s}$ is the relativistic factor of the shock wave.
This is a way beyond the simple thermal shock wave acceleration. An
intense electrostatic field moving at relativistic velocity is required.
One possibility might be the plasma wakefield. It is works very good
for electron acceleration, however it may be not intense enough to
trap protons, although there are some proposals to trap pre-accelerated
protons and accelerate them further by the wakefield to very high
energes \cite{shen2009}. 

We propose here that the electrostatic field of LPA can serve as the
perfect relativistic piston. Its velocity is close to the light speed
and the charge separation field 
is sufficiently high to trap background protons. In this paper,
we develope an analytical model and use particle-in-cell (PIC) simulations
to show that when a thin foil driven by ultra-intense circularly polarized
(CP) laser pulses, with peak amplitude $a_{0}=eE_{L}/m_{e}\omega_{0}c$
from 50 to 200, passes through an underdense  plasma region,
reflected protons can achieve energes up to hundreds of
GeV. Here $e$ and $m_{e}$ are fundamental charge and electron mass,
$E{}_{L}$ and $\omega{}_{0}$ are electric field amplitude and angular
frequency of the laser pulse, $c$ is the light speed, respectively. 

The expression of the accelerating field amplitude as a function of
time is derived, which, together with the foil momentum equation,
well describes the movement of the background protons and the trapping
condition. The scaling laws indicate that the peak energy increases
with $t^{2/3}$ and $a_{0}^{2}$. A proof-of-principle two-dimensional
(2D) simulation is also performed to show the validity of our analysis
in a  multi-dimensional geometry.

\begin{figure}
\includegraphics[width=0.9\columnwidth]{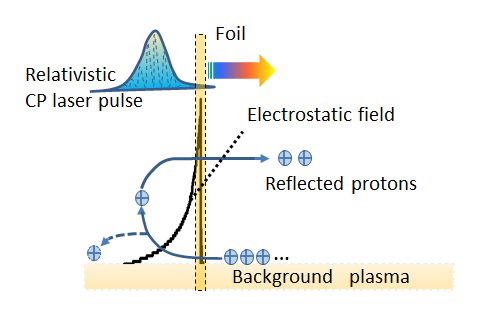}

\caption{\label{fig:Figure1} {\small The electrostatic field driven by a relativistic
CP laser propagates in the underdense background
plasma, traps background protons and reflects them to ultra-high energy. }}
\end{figure}

The mechanism is sketched in Fig. \ref{fig:Figure1} . As known in
LPA, the laser-driven foil is
accompanied by an intense charge separation field. When it propagates
through the low density plasma, background protons can be continuously
trapped in certain conditions. The foil plays the role of a relativistic piston. In the frame
of the relativistic piston, background protons clash towards the electrostatic
field at the foil velocity, then are slowed down by the field in and
behind the foil and finally reflected.

In LPA, protons are accelerated by the field
localized in the ultra-thin skin layer only, leaving the enormous
"dragging fiel"' (DF) unused. In our scheme, the background protons
pass through the skin layer and are greatly accelerated by the DF,
which contains much more electrostatic potential than the field inside
the layer. As a result, the reflected protons can obtain
energy a magnitude higher than foil protons in the simple LPA regime. 

\begin{figure}
\subfigure{\includegraphics[width=0.48\columnwidth]{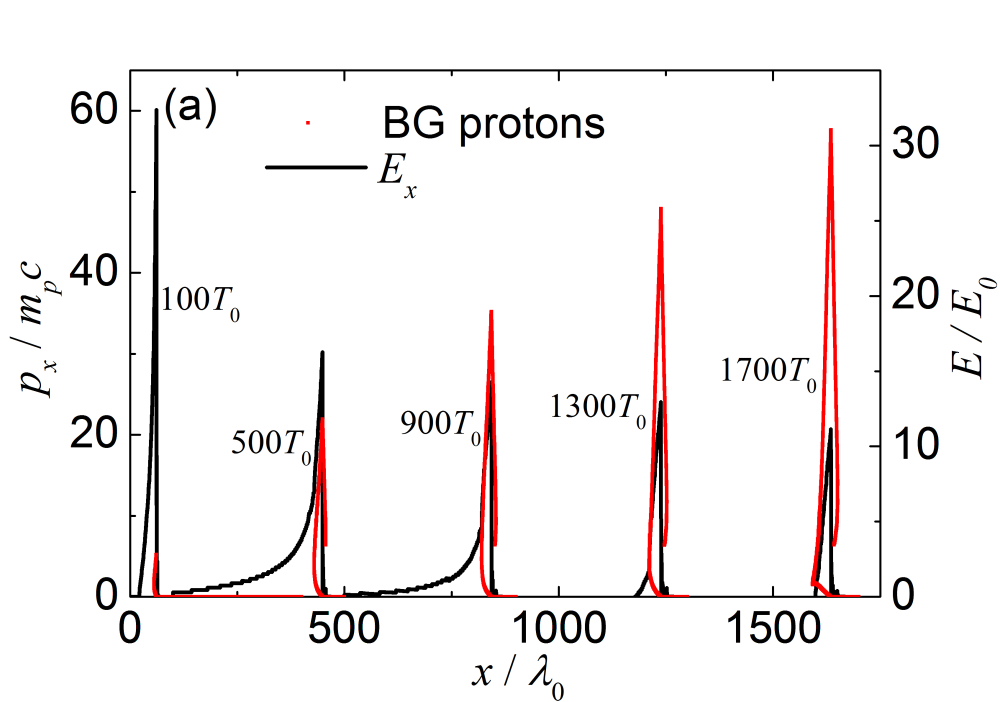}}
\subfigure{\includegraphics[width=0.48\columnwidth]{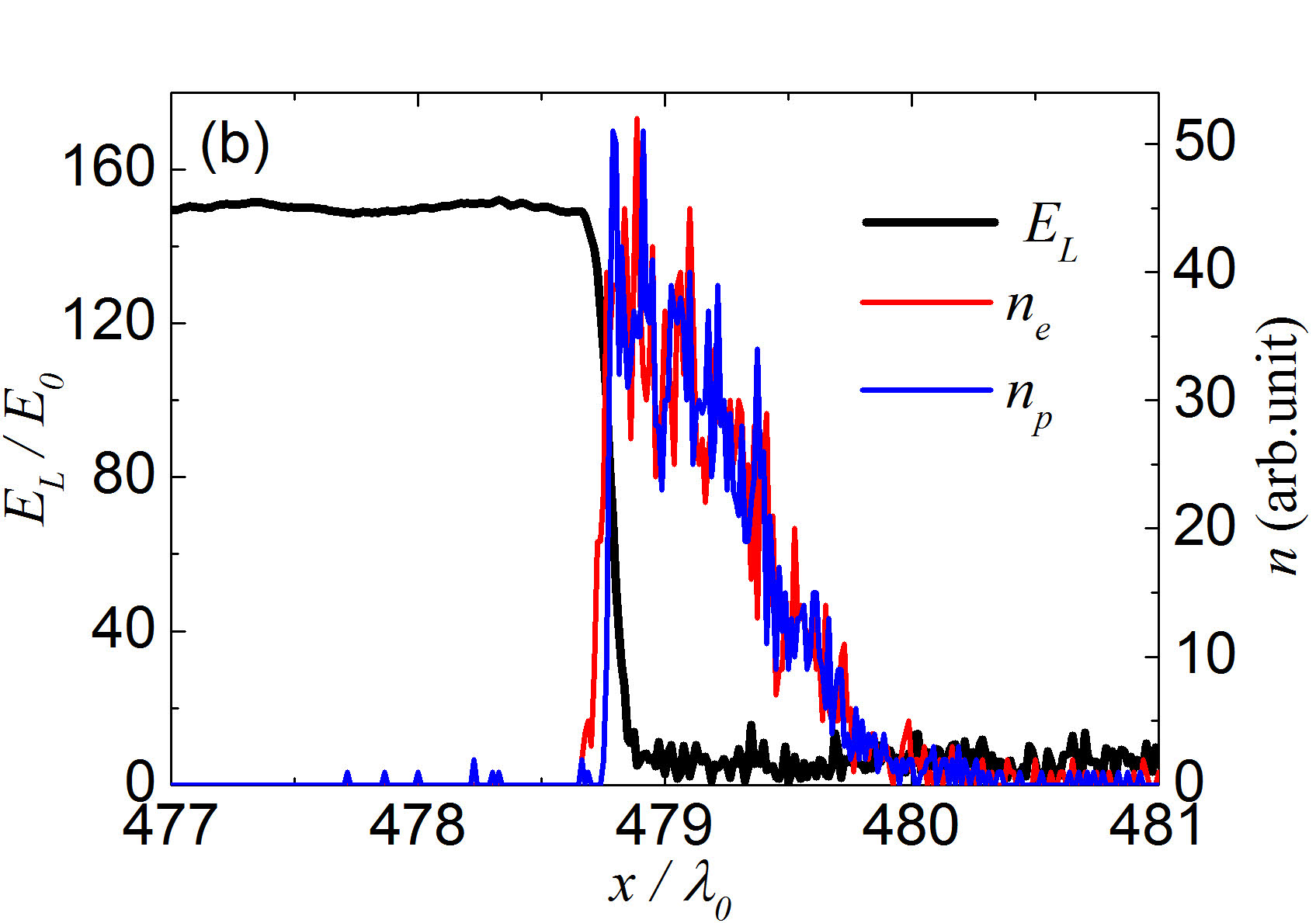}}

\noindent \subfigure{\includegraphics[width=0.48\columnwidth]{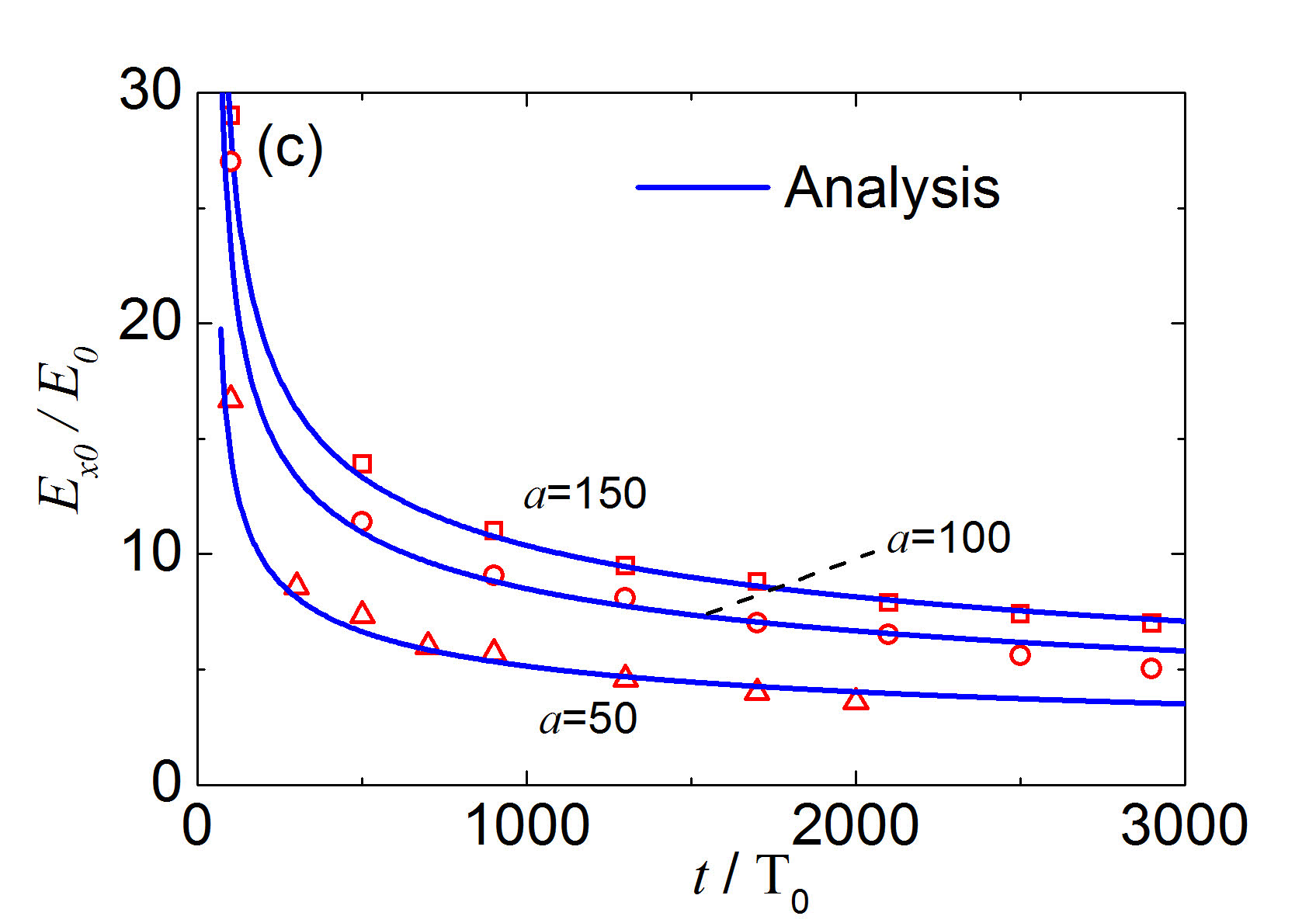}}
\subfigure{\includegraphics[width=0.48\columnwidth]{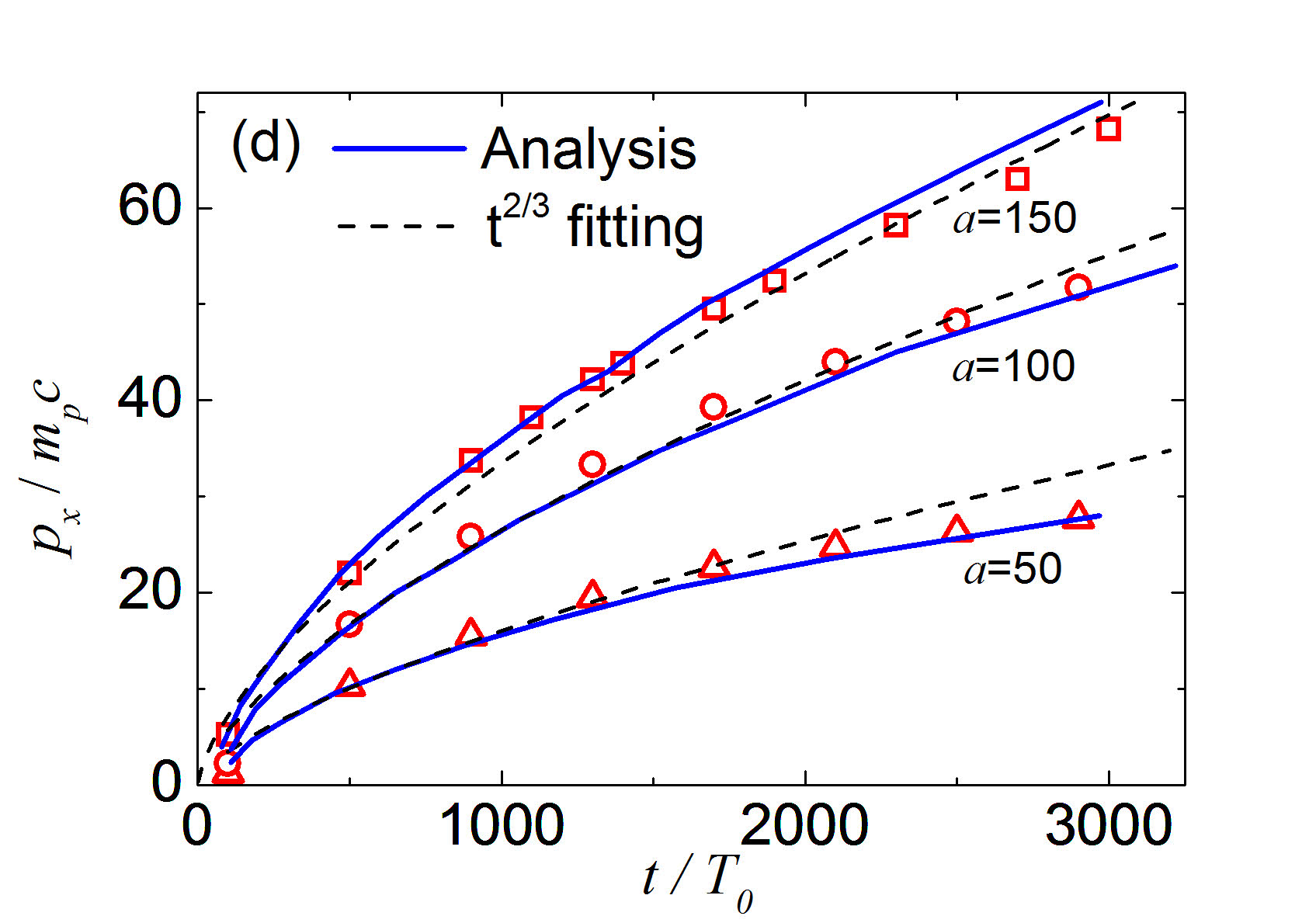}}

\caption{\label{fig:Fig. 2} {\small a) Distributions of electrostatic field
(blue solid) and momentum of background (BG) protons (red dotted)
at different stages with $a{}_{0}=150$, $n{}_{f}=150n_{c}$, $d{}_{f}=0.57\lambda_{0}$
and $n{}_{b}=0.001n_{c}$. b) Distributions of laser
field (black solid), electron density (red solid) and proton density
(blue solid) at $t=500T_{0}$. c) Evolution of the peak electrostatic
fields of $a{}_{0}=50$, 100 and 150, from analytical model (blue
solid) and one-dimensional simulations. d) Evolution of the maximum
energy of background protons with $a{}_{0}=50$, 100 and 150, from
analytical model (blue solid) and simulations. Here the dashed lines
show that the peak energy increases with $\sim t^{2/3}$.}}
\end{figure}

To demonstrate the mechanism a 1D PIC simulation is performed by the
VLPL code \cite{pukhov1999}. A dense thin foil, with density of $n{}_{f}=150n_{c}$
and thickness of $d{}_{f}=0.57\lambda_{0}$, is driven by a 
CP laser pulse with peak amplitude $a{}_{0}=150$, where $n{}_{c}=m_{e}\omega_{0}^{2}/4\pi e^{2}$
is the critical density and $\lambda{}_{0}=0.8\mu m$ is the laser
wavelength. For simplicity, we used a trapezoidal laser pulse
. Its amplitude increased linearly to the maximum in $2T{}_{0}$
and then stayed constant ($T{}_{0}$ is the laser period). Here the
foil thickness is optimized for LPA \cite{Wang2011}. A
background plasma with density of $n{}_{b}=0.001n_{c}$ is located
behind the foil. Fig. \ref{fig:Fig. 2}(a) shows distributions of
the electrostatic field and momentum of background protons. The field
decays quickly at first and much more slowly later.
It is so intense, well beyond $E{}_{0}=m_{e}\omega_{0}c/e$
even at the end of the interaction, that background protons are gradually trapped and reflected to about 70 GeV in 3mm, while
protons in the front foil have the peak energy below 10 GeV. 

Trapping and accelerating of charged particles are determined by the
velocity, amplitude and length scale of the DF structure. Since the
DF co-moves with the foil, its dynamics can be derived by the momentum
equation of the piston \cite{esirkepov2004} 
\begin{equation}
\frac{d(\gamma_{f}\beta_{f})}{dt}=\frac{m_{e}}{m_{i}}\frac{2a_{(t-x_{f})}^{2}}{N_{f}D_{f}}\frac{1-\beta_{f}}{1+\beta_{f}}.\label{eq:PistonDynamics}
\end{equation}

\noindent here $\beta{}_{f}$ is the foil velocity normalized by $c$, $\gamma{}_{f}=(1-\beta_{f}^{2})^{-1/2}$, and $m{}_{i}$
is the ion mass. $N{}_{f}$ and $D{}_{f}$ are initial density and
thickness of the foil normalized by the ciritical density $n_{c}$
and laser wavelength $\lambda{}_{0}$. All lengths and the time are
normalized by $\lambda{}_{0}$ and $T{}_{0}$, respectively. The velocity
of the DF structure is thus $\beta{}_{E}$ =$\beta{}_{f}$, and can
be obtained by solving Eq. \eqref{eq:PistonDynamics}.

More critical is the evolution of field amplitude. Former researches
suggested that considering the balance of the electron thin layer
between the light pressure and electrostatic force, the peak amplitude
of the DF should be proportional to the Doppler factor $\left(1-\beta_{f}\right)/\left(1+\beta_{f}\right)$
\cite{Macchi2010}. This means that it would drop by $1/4\gamma{}_{f}^{2}$
in the relativistic regime. In the above simulation $\gamma{}_{f}\thickapprox5$
at $t=500T_{0}$, and the DF amplitude should decrease by some factor
100. However, as seen in Fig. \ref{fig:Fig. 2}(a) The DF at $t=500T_{0}$ is still above $E/E_{0}>15$, roughly one tenth of initial maximum field. A new scaling should then be deduced. In Fig. \ref{fig:Fig. 2}(b) the
distributions of laser field and electrons at $t=500T_{0}$ are presented.
The laser ponderomotive force acts on only a small portion
of the electrons at the surface and not on the whole electron layer.
This is because the light pressure is severely weakened when the foil
becomes relativistic so that it cannot confine all foil electrons. The situation is then similar to the electrostatic shock or
the hole-boring process \cite{macchi2005}, except the whole system
is moving fast. 

Since the foil velocity varies slowly with time, it is reasonable
to assume the piston frame as an inertial reference system,
where relationships in Ref. \cite{macchi2005} can be applied after
Lorentz transforming all quantities. Considering the balance of the
electron skin layer instead of the whole electron layer, the peak
DF is $E{}_{x0}^{'}\sim\sqrt{P_{rad}^{'}}$ , where $P{}_{rad}^{'}$
is the light pressure in the moving frame $K^{'}$. The longitudinal
electric field and light pressure turn out to be Lorentz invariants,
thus the peak amplitude in laboratory frame is conveniently obtained
by $E{}_{x0}\sim\sqrt{P_{rad}}\sim\sqrt{(1-\beta_{f})/(1+\beta_{f})}$.
On the other hand, the ponderomotive force acting on electrons is
approximately $|e\mathbf{\boldsymbol{\beta}}_{e}^{'}\times\mathbf{B}_{L}^{'}|\approx eE_{L}^{'}$
, therefore $E{}_{x0}^{'}\sim E_{L}^{'}$, giving the normalized peak
amplitude of

\begin{equation}
E_{x0}(t)\approx a\sqrt{\frac{1-\beta_{f}}{1+\beta_{f}}}.\label{eq:Ex}
\end{equation}

\noindent The both estimations above show that the field 
scales proportional to the square root of the Doppler factor, i.e.,
about $1/2\gamma{}_{f}$ . This is a much weaker decay than
formerly assumed. We compare
the results from Eqs. \eqref{eq:PistonDynamics} and \eqref{eq:Ex}h  with 1D PIC simulation results in Fig. \ref{fig:Fig. 2}.
One can see that for $a$=50, 100 and 150, Eq. \ref{eq:Ex} describes
the simulation results properly. This new scaling is very important
to guarantee the DF structure can be maintained for a long distance
and does not vanish in relativistic regime.

The
dynamics of background plasma protons encountered in the field of
the piston are then derived by

\noindent 
\begin{equation}
\frac{d(\gamma_{b}\beta_{b})}{dt}=\frac{2\pi m_{e}E_{x}(t,x_{b})}{m_{p}}.\label{eq:BG protons}
\end{equation}

\noindent As seen from Fig. \ref{fig:Figure1} and Fig. \ref{fig:Fig. 2}(a),
the DF drops almost linearly with the distance shortly behind the
foil.The long weak tail far behind the foil is neglected since protons located there are no longer
trapped. We simplify the description, by assuming the charge separation
field behind the skin layer decays to zero in a length scale of $d{}_{E}$

\noindent 
\begin{equation}
E_{x}(t,x_{b})=\begin{cases}
E_{x0}(t)(1-\frac{x_{f}-x_{b}}{d_{E}}), & x_{f}-d_{E}\leq x_{b}\leq x_{f}\\
0, & elsewhere
\end{cases}.\label{eq:DF}
\end{equation}

\noindent Here the $x{}_{b}$, $\beta{}_{b}$, $\gamma{}_{b}$ are
position, normalized velocity and $\gamma-$factor of the background
protons. The scale length of the electrostatic field localized inside
the skin layer is so small that we ignore it.
Simulations suggest that the DF scale is about $d_{E}\approx40\mu\textrm{m}$.
Protons initially located at different positions of $x{}_{b}|_{t=0}$,
which are also the injected positions, obtain final energy when $x{}_{b}=x_{f}$.
So the maximum energy of all reflected protons as a function of time
can be analytically derived by changing the $x{}_{b}|_{t=0}$ in Eqs.
\eqref{eq:BG protons} and \eqref{eq:DF}.

Eqs. \eqref{eq:PistonDynamics}-\eqref{eq:DF} offer a complete description
of the mechanism. Peak momenta of the trapped protons at different
simulation times and from analytical model are shown in Fig. \ref{fig:Fig. 2}(d),
for $a$=50, 100 and 150, respectively. Again, the analytical results
and the simulations are in a good agreement. Simulation data is also
fitted, showing that the peak momentum increases with time according
to the power law $\propto t^{2/3}$ , much faster than in the simple
LPA ($\propto t^{1/3}$) \cite{esirkepov2004}. The great difference
is due to the fact that the light pressure - that drives the foil
- decays as $1/4\gamma{}_{f}^{2}$ while the DF accelerating the background
protons decays much slower, as $1/2\gamma_{f}$ . Within a few millimeters
the protons are accelerated close to 100GeV,
while in LPA it is nearly impossible to produce protons beyond ten
GeV even in 1D simulations.

\begin{figure}
\begin{centering}
\includegraphics[width=0.9\columnwidth]{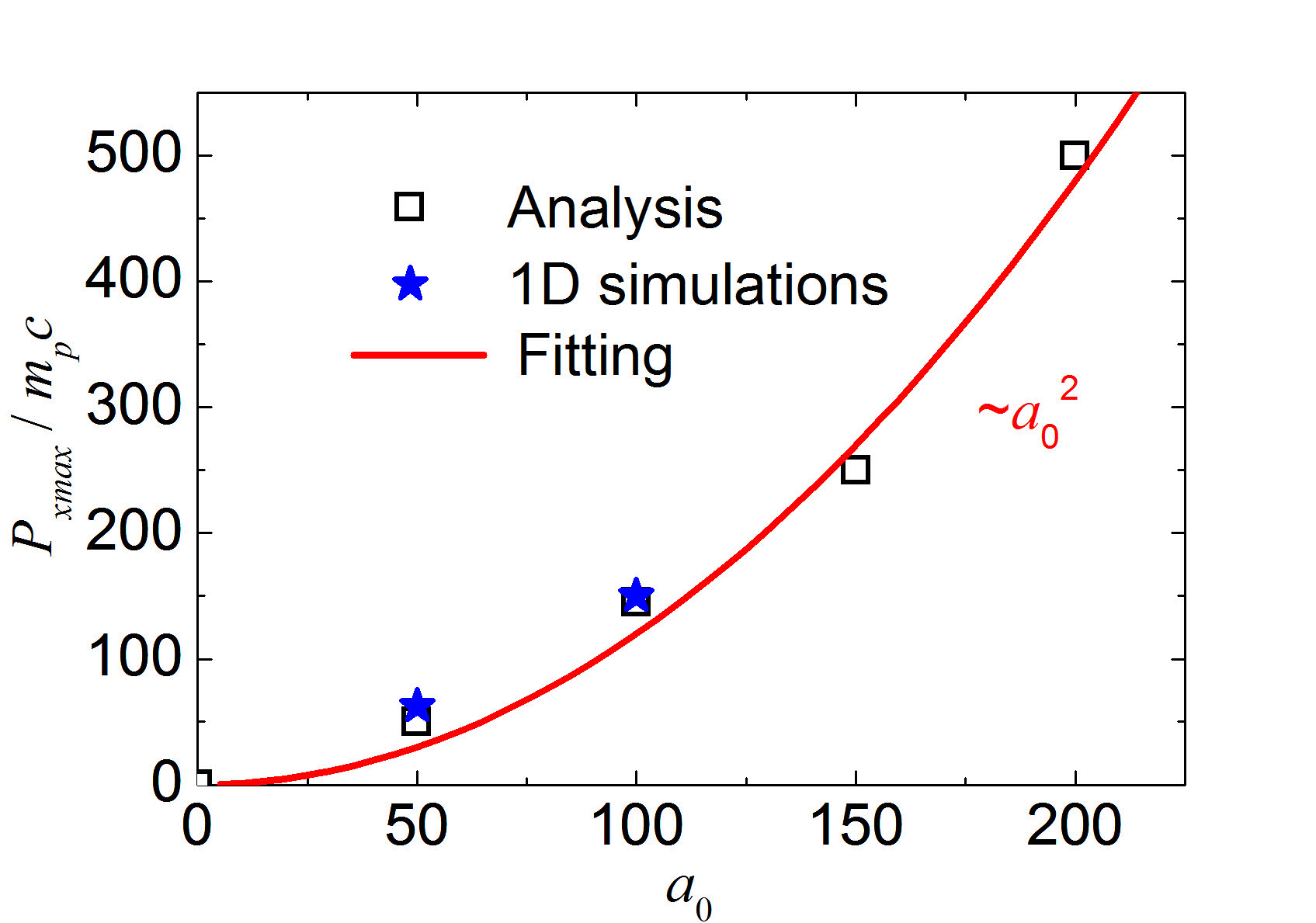}
\par\end{centering}

\caption{\label{fig:Fig.3} {\small Scaling law of the maximum achievable
momentum versus peak laser amplitude from analytical model (black
square) and simulations (blue pentacle). The fitted red line indicates
the power-law of $\sim a{}_{0}^{2}$. }}
\end{figure}

In DF acceleration, the piston is continuously accelerated. Thus,
the later local protons are injected, the higher energy they
obtain. However, the DF amplitude also decays
meanwhile so that after certain point it becomes not intense
enough to trap stationary protons. The trapping condition is obtained
by balancing kinetic energy of the incoming protons and the DF electrostatic
potential in the relativistic piston frame

\noindent 
\begin{equation}
\frac{1}{2}eE_{x0}^{'}(t)d_{E}^{'}=(\gamma_{f}-1)m_{p}c^{2}.\label{eq:Balance}
\end{equation}

\noindent where $E{}_{x0}^{'}(t)=E_{x0}(t)$ and $d{}_{E}^{'}=\gamma_{f}d_{E}$
is the peak amplitude and length scale of the DG field in the moving
frame. The critical foil $\gamma-$factor beyond which background
protons are no longer trapped is then 

\noindent 
\begin{equation}
\gamma_{fc}\approx1+\frac{\pi m_{e}}{m_{p}}\frac{a_{0}D_{E}}{2}.\label{eq:CriticalGammaFoil}
\end{equation}

\noindent For $a{}_{0}=150$ and $D_{E}=d{}_{E}/\lambda_{0}=40$,
the critical value is about $\gamma{}_{fc}\approx6$ . After the foil’s
$\gamma-$factor reaches the critical value, latish protons would
pass through the DF structure without being trapped any more. This defines the
maximum energy obtainable by this mechanism. In Fig. \ref{fig:Fig.3}
the scaling law between the maximum obtainable energy and $a{}_{0}$ from simulations is compared with the  model. The simulation results are only shown for $a{}_{0}=50$
and 100. For higher amplitudes, it would take too much
 time to reach the final stage (about $10^{4}$
μm). Fig. \ref{fig:Fig.3} indicates that the maximum energy is proportional
to the square of the laser amplitude $\sim a_{0}^{2}$. This mechanism
allows to accelerate protons to hundreds of GeV. 

In 1D simulations we chose very low background
densities to ensure that the DF is not affected. In the multi-dimensional
geometry, the situation becomes more complicated. Firstly, instabilities
may destroy the
thin foil during the interaction. This could be restrained
by using a highly relativistic laser pulse, e.g., $a{}_{0}>100$ in
our case. Secondly, the foil deformation due
to the laser intensity distribution will cause diffraction.
Reflected parts of the laser fields interfere with each other
behind the moving piston. The DF structure is well formed at the beginning,
however when the perturbation arrives at the axis, it may breakdown
before any protons being trapped.

\begin{figure}
\begin{centering}
\subfigure{\includegraphics[width=0.48\columnwidth]{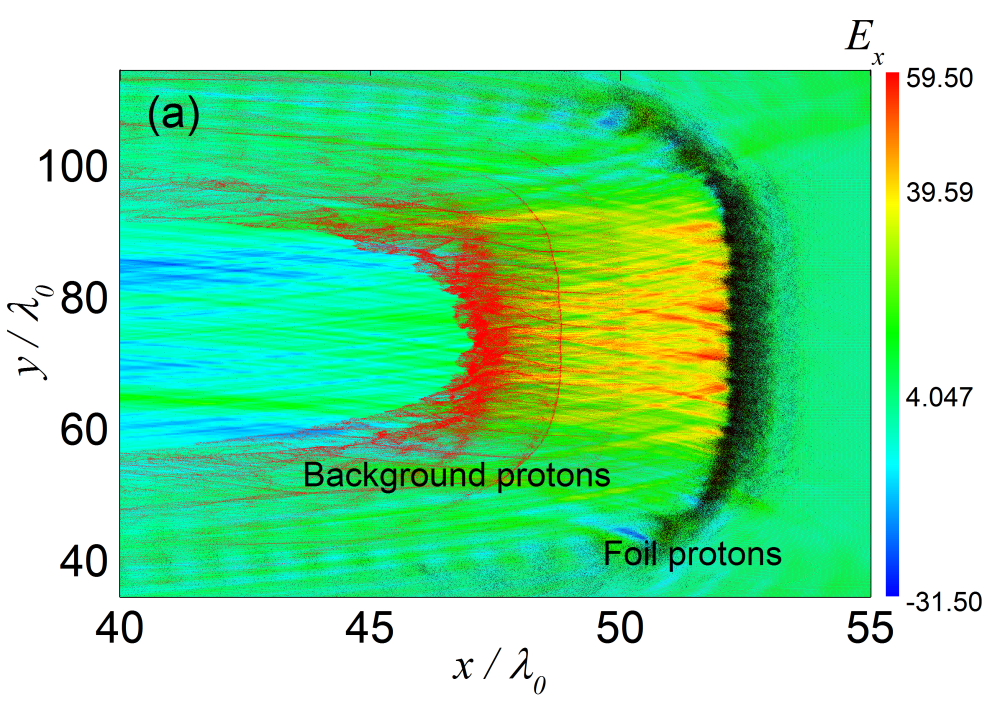}}
\subfigure{\includegraphics[width=0.48\columnwidth]{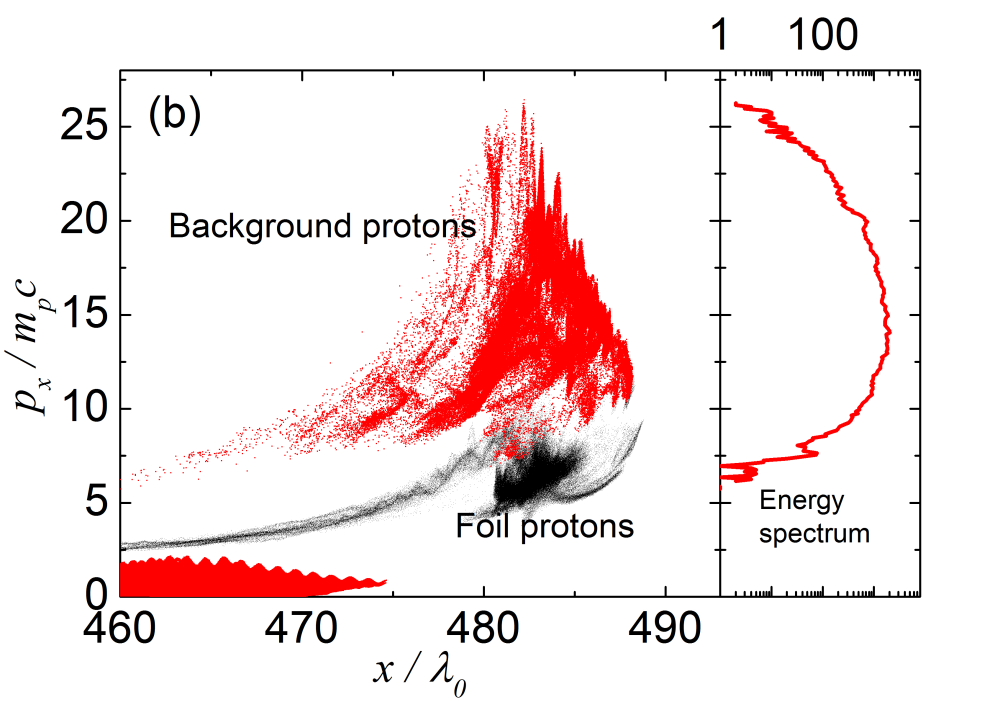}}
\par\end{centering}
\caption{\label{fig:Fig.4} {\small Two-dimensional results with $a{}_{0}=200$,
$n{}_{f}=80n_{c}$, $d{}_{f}=0.62\lambda_{0}$ and $n{}_{b}=0.1n_{c}$.
a) Space distributions of the foil protons (black dotted), background
protons (red dotted) and the electrostatic field at $t=60T_{0}$;
b) Momentum distributions of the foil protons (black dotted) and background
protons (red dotted). The right panel is the energy spectrum of background
protons.}}
\end{figure}

To address this, we increase the background plasma density.
Background electrons are snow-plowed by
the driving laser and pile up inside the piston. As a result, higher
background density could induce two effects. Firstly, the relativistic piston is slowed down. Secondly,
the DF might be enhanced by the additional charge
separating field between piled-up electrons and left-behind protons.
The higher background density easies the trapping of local protons.The 2D simulation results are shown
in Fig. \ref{fig:Fig.4}. The laser pulse has the same trapezoidal
profile as in 1D simulations in time domain and a super-Gaussian
distribution transversely \textasciitilde{}$e{}^{-(y/w_{y})^{4}}$
, where $w{}_{y}=35\mu m$. To restrain instabilities,
the laser peak amplitude is increased to $a{}_{0}=200$. The diffraction
and reflection of the incident pulse is minimized by using density
matching foil, i.e., the foil is with density distribution of $n{}_{f}(y)=n_{0}e^{-(y/w_{y})^{4}}$\cite{Chen2009}.
The foil density and thickness are $n{}_{f}=80n_{c}$ and $d{}_{f}=0.62\lambda_{0}$
while the background plasma density is increased to $n{}_{b}=0.1n_{c}$.

Fig. \ref{fig:Fig.4}(a) shows distributions of the DF, electrons
in the foil and background protons at $t=60T_{0}$. It is seen that
the DF is well formed behind the thin layer. Before the instabilities
could develop perturbing the accelerating field, some of the background
protons are trapped and accelerated. As seen in Fig. \ref{fig:Fig.4}(b),
the maximum energy of reflected protons at $t=500T_{0}$ is about
25GeV, approximately 4 times the peak proton energy in LPA. Some foil protons left behind the layer are also trapped and
accelerated to high energy. The total energy of reflected background
protons is about 3\% of all foil protons, which is quite considerable
concerning the low background density. One would expect higher energy
efficiency by using denser background plasma.

In conclusion, a method to laser accelerate protons beyond tens of
GeV is proposed. It takes advantage of the electrostatic field dragging
behind the flying foil in LPA to trap background protons. The dragging
field keeps trapping and accelerating protons to energies greatly
higher than that obtained in LPA. The analytical model shows that
the electrostatic field decays more slowly than formerly assumed,
resulting in larger energy increasing rate ($\sim t^{2/3}$).The final maximum
energy scales as $\sim a{}_{0}^{2}$. 2D simulations proved the proposal, generating
protons with peak energy of about 25GeV. 

This work is supported by a fellowship of the Alexander von Humboldt
Foundation for Liangliang Ji, National Natural Science Foundation of China (Project No. 11125526, and No. 60921004)

\end{document}